\documentclass[11pt]{article}
\usepackage{amssymb}
\usepackage{amsmath}
\usepackage[onehalfspacing]{setspace}
\usepackage{geometry}
\usepackage{natbib}

\newtheorem{theorem}{Theorem}

\newtheorem{lemma}[theorem]{Lemma}

\newtheorem{proposition}[theorem]{Proposition}
\newtheorem{remark}[theorem]{Remark}

\input{tcilatex}
\begin{document}

\begin{center}
\mathstrut

{\LARGE Competition for Budget-Constrained Buyers: Exploring All-Pay Auctions%
}

\mathstrut

\begin{equation*}
\begin{tabular}{c}
Cemil Selcuk \\ 
Cardiff Business School, Cardiff University \\ 
Colum Drive, Cardiff, UK \\ 
selcukc@cardiff.ac.uk, +44 (0)29 2087 0831 \\ 
\end{tabular}%
\end{equation*}

\bigskip

\bigskip
\end{center}

\section{Introduction}

This note pursues two primary objectives. First, we analyze the outcomes of
an all-pay auction within a store where buyers with and without financial
constraints arrive at varying rates, and where buyer types are private
information. Second, we investigate the selection of an auction format
(comprising first-price, second-price, and all-pay formats) in a competitive
search setting, where sellers try to attract customers. Our results indicate
that if the budget constraint is not too restrictive, the all-pay rule
emerges as the preferred selling format in the unique symmetric equilibrium.
This is thanks to its ability to prompt buyers to submit lower bids, thereby
generally avoiding the budget constraints, while allowing the seller to
collect all bids.

\section{Model}

\emph{Environment. }The economy consists of a large number of risk-neutral
buyers and sellers, with an overall buyer-seller ratio denoted by $\lambda $%
. Each seller has one unit of a good and aims to sell it at a price
exceeding his reservation price, which is zero. Similarly, each buyer seeks
to purchase one unit and is willing to pay up to their reservation price,
set at one. While buyers share identical valuations of the good, their
purchasing abilities vary. Specifically, a fraction $\sigma $ of buyers,
known as "low types", have constrained budgets and can pay only up to $b<1$,
while the remaining buyers, termed "high types", can pay up to 1. The type
of buyer is private information, but the parameters $\lambda $, $\sigma $,
and $b$ are common knowledge.

The game proceeds over the course of two stages. In the first stage sellers
simultaneously and independently choose an auction format $m$ and a reserve
price $r_{m}.$ The set of formats consists of first-price auction,
second-price auction and all-pay auction. In the second stage buyers observe
sellers' selections and select one store to visit. If the customer is alone
at the store then he pays the reserve price and obtains the good. If there
are $n\geq 2$ buyers then bidding ensues and the winner as well as the sale
price are determined based on the auction format. If trade takes place at
price $r$ then the seller realizes payoff $r$, the buyer realizes $1-r$
whereas those who do not trade earn zero.

\emph{Demand Distribution.} Following the competitive search literature, we
focus on visiting strategies that are symmetric and anonymous on and off the
equilibrium path, which, in a large market, imply that the distribution of
demand at any store is Poisson 
\citep{burdett_pricing_2001,
shimer_assignment_2005}. Therefore, the probability that a seller with the
terms $\left( m,r_{m}\right) $ meets $n$ customers of type $i=h,l$ is given
by $z_{n}\left( x_{i,m}\right) $ where 
\begin{equation}
z_{n}\left( x\right) :=\frac{e^{-x}x^{n}}{n!}\text{.}  \label{z(n)}
\end{equation}%
Since high types and low types arrive at independent Poisson rates $x_{h,m}$
and $x_{l,m},$ the distribution of the \emph{total demand }is also Poisson
with $x_{h,m}+x_{l,m}$. Both $x_{h,m}$ and $x_{l,m}$ are endogenous and they
depend on what the seller posts and how it compares with the rest of the
market.

\section{Auctions}

Let $u_{i,m}\left( n\right) $ denote the expected utility of a type $i$
buyer at a store that trades via rule $m$ and has $n$ customers, including
the buyer himself. Similarly let $\pi _{m}\left( n\right) $ denote a store's
expected profit conditional on trading via rule $m$ and having $n$
customers. Below we pin down these payoffs for all auction rules. (We drop
the subscript $m$ when understood.) Bidding ensues if $n\geq 2,$ so consider
a store with $n\geq 2$ customers. Low budget types arrive at Poisson rate $%
x_{l}$ and high budget types arrive at rate $x_{h}$. The distribution of the
number of low types, therefore, is $binomial(n,\theta )$ where%
\begin{equation*}
\theta =\frac{x_{l}}{x_{h}+x_{l}}.
\end{equation*}%
Note that $\theta $ is the probability that a buyer is a low type and it is
endogenous as it depends on the endogenous arrival rates $x_{h}$ and $x_{l}$.

\begin{remark}
With a first-price auction, there exits a unique symmetric equilibrium in
which low type buyers bid $b$ while high type buyers randomize within $\left[
b,1-\left( 1-b\right) \theta ^{n-1}\right] $. Similarly, with a second price
auction in the unique symmetric equilibrium low type buyers bid $b$ while
high types bid their valuation, $1$. Under both auction formats buyers'
expected earnings are%
\begin{equation*}
u_{h}\left( n\right) =\theta ^{n-1}\left( 1-b\right) \ \ \ \ \ \text{and}\ \
\ \ u_{l}\left( n\right) =\frac{\theta ^{n-1}}{n}\left( 1-b\right) \text{
for }n\geq 2
\end{equation*}%
whereas the seller earns%
\begin{equation}
\pi \left( n\right) =\left[ \theta ^{n}+n\theta ^{n-1}\left( 1-\theta
\right) \right] b+1-\theta ^{n}-n\theta ^{n-1}\left( 1-\theta \right) \text{
for }n\geq 2.  \label{pi(n) - 1st 2nd}
\end{equation}
\end{remark}

The remark is based on \cite{SelcukGEB}. In both auction formats, low types
bid their budget $b$. Conversely, high types exhibit less aggressive bidding
in the first-price auction compared to the second-price auction. However,
their expected earnings remain equal. Moreover, both auction formats yield
identical expected revenue for the seller. Given this payoff equivalence, in
a competitive setting sellers and buyers are indifferent to adopting or
joining either auction format. From now on we will treat first and second
price auctions equally, occasionally referring to them as \emph{standard }%
auctions.

\begin{proposition}
\label{Proposition-AuctionOutcome-ALLPAY}All-pay auction. If $b<\theta
^{n-1} $ then a unique symmetric equilibrium exits where high types
randomize in the interval $[b,1-\theta ^{n-1}+b]$ according to cdf%
\begin{equation*}
G_{h}\left( p\right) =\frac{\left( p+\theta ^{n-1}-b\right) ^{\frac{1}{n-1}%
}-\theta }{1-\theta }.
\end{equation*}%
For low types there are two scenarios. If $b<\frac{\theta ^{n-1}}{n}$ then
they all bid $b$, but if $\frac{\theta ^{n-1}}{n}\leq b<\theta ^{n-1}$ then
they employ a strategy $G_{l}\left( p\right) $ with support $[0,b]$ that has
an atom at $b.$ The size of the atom $\mu $ falls as $b$ rises$.$
Specifically, we have%
\begin{equation*}
G_{l}\left( p\right) =\left\{ 
\begin{array}{ccc}
\frac{p^{\frac{1}{n-1}}}{\theta } & \text{if} & 0\leq p<\left( 1-\mu \right)
^{n-1}\theta ^{n-1} \\ 
1-\mu & \text{if} & \left( 1-\mu \right) ^{n-1}\theta ^{n-1}\leq p<b \\ 
1 & \text{if} & p=b%
\end{array}%
\right. .
\end{equation*}%
In this parameter region equilibrium payoffs are given by%
\begin{equation*}
u_{h}\left( n\right) =\theta ^{n-1}-b,\text{ while }u_{l}\left( n\right) =0%
\text{ if }b\geq \theta ^{n-1}/n\text{ \ and \ }u_{l}\left( n\right) =\frac{%
\theta ^{n-1}}{n}-b\text{ otherwise}.
\end{equation*}

If $\theta ^{n-1}\leq b$ then in any symmetric equilibrium the seller
extracts the entire surplus while all bidders earn zero i.e. $\pi \left(
n\right) =1,$ and $u_{h}\left( n\right) =u_{l}\left( n\right) =0.$ The
following strategies constitute such an outcome:\ low types continuously
randomize in $[0,\theta ^{n-1}]$ whereas high types continuously randomize
in $[\theta ^{n-1},1]$ according to density functions%
\begin{equation}
G_{l}\left( p\right) =\frac{p^{\frac{1}{n-1}}}{\theta }\text{ \ \ \ and \ \ }%
G_{h}\left( p\right) =\frac{p^{\frac{1}{n-1}}-\theta }{1-\theta }.
\label{GL-GH-ZeroPayoffs}
\end{equation}
\end{proposition}

In the all-pay format, participants are obliged to pay their bid regardless
of whether they win or lose. As a result, compared to first or second price
auctions, buyers bid smaller amounts. Bid amounts decrease particularly
among low types. Whereas in first and second price auctions, low types would
bid their entire budget $b$, they now engage in randomization within a lower
range. High types also reduce their bids, though the effect on them is less
pronounced.

Despite the lower bids, the seller manages to collect a higher amount of
revenue than first or second price auctions. For instance when $b\geq \theta
^{n-1}$, i.e. when the budget is not too small, we see that all-pay auctions
yield $\pi \left( n\right) =1$ whereas first and second price auctions yield
a strictly smaller $\pi \left( n\right) $. Again, this is owing to the
auction format allowing the seller to collect all the bids. Absent budget
constraints, revenue equivalence among fixed-price, second-price and all-pay
auctions is well known in the literature. We show that with financially
constrained buyers, the all-pay format revenue-dominates the first or second
price formats. \cite{che_expected_1996} prove a similar result when the
budget distribution is continuous. We show that the result remains valid
under a discrete distribution. This advantage will play a key role when
sellers pick an auction format to compete for buyers.

\begin{equation*}
\text{Figure of cdfs here}
\end{equation*}

The amount of bids depend on $\theta ,$ the percentage of low types expected
to be present in the auction. This parameter is endogenous and critically
impacts the auction process. For instance suppose a seller attracts high
types only. Then $\theta =0,$ and per the Proposition, and they randomize in
the interval $\left[ 0,1\right] $ according to $G=p^{\frac{1}{n-1}}.$ \cite%
{baye_all-pay_1996} obtain this result in a setting with no budget
constraints (see Theorem 1 therein).

\section{Competitive Search}

Let $s$ represent standard (first or second price) formats and $ap$
represent the all pay format. A type $i$ buyer's expected utility from
visiting a store competing with rule $m=s,$ $ap$ is given by%
\begin{equation}
U_{i,m}\left( r_{m},x_{h,m},x_{l,m}\right) =\sum_{n=0}^{\infty }z_{n}\left(
x_{h,m}+x_{l,m}\right) u_{i,m}\left( n+1\right) ,  \label{Uim - RAw}
\end{equation}%
With probability $z_{n}\left( \cdot \right) $ the buyer finds $n=0,1,..$ 
\emph{other} customers at the same store; so, in total there are $n+1$
customers (including himself) and the expected utility corresponding to this
scenario is $u_{i,m}\left( n+1\right) .$ Now consider a seller who competes
with rule $m.$ His expected profit is given by%
\begin{equation*}
\Pi _{m}\left( r_{m},x_{h,m},x_{l,m}\right) =\sum_{n=1}^{\infty }z_{n}\left(
x_{h,m}+x_{l,m}\right) \pi _{m}\left( n\right) .
\end{equation*}
With probability $z_{n}\left( \cdot \right) $ the seller gets $n$ customers
and the corresponding payoff associated with this scenario is $\pi
_{m}\left( n\right) .$ The following Lemma links $\Pi $ to the $U_{i}$s$.$

\begin{lemma}
\label{Lemma - Pi and U} The following relationship holds both for first and
second price formats as well as for the all-pay format:%
\begin{equation}
\Pi _{m}=1-z_{0}\left( x_{h,m}+x_{l,m}\right) -x_{h,m}U_{h,m}-x_{l,m}U_{l,m},%
\text{ where }m=s,ap.  \label{kocum be}
\end{equation}
\end{lemma}

\noindent The expression $1-z_{0}\left( x_{h,m}+x_{l,m}\right) $ can be
interpreted as the expected revenue. It is the value created by a sale
(one), multiplied by the probability of trading. The expression $%
x_{h,m}U_{h,m}+x_{l,m}U_{l,m}$ can be interpreted as the expected cost. The
seller promises a payoff $U_{h,m}$ to each high type and $U_{l,m}$ to each
low type customer. On average he gets $x_{h,m}$ high type and $x_{l.m}$ low
type customers; so the total cost equals to $x_{h,m}U_{h,m}+x_{l,m}U_{l,m}$.
The profit $\Pi _{m}$ is simply the difference between the revenue and the
cost.

Following the competitive search, let $\Omega _{i}$ denote the maximum
expected utility\ ("market utility") a type $i$ customer can obtain in the
entire market.\footnote{%
The market utility approach greatly facilitates the characterization of
equilibrium and, therefore, is standard in the directed search literature.
For an extended discussion see \cite{galenianos_game-theoretic_2012}.} For
now we treat $\Omega _{i}$ as given, subsequently it will be determined
endogenously. Consider an individual seller who advertises $\left(
m,r_{m}\right) $ and suppose that high and low type buyers respond to this
advertisement with arrival rates $x_{h,m}\geq 0$ and $x_{l,m}\geq 0.$ These
rates satisfy%
\begin{equation}
x_{i,m}\text{ }\left\{ 
\begin{array}{lcc}
>0 & \text{if} & U_{i,m}\left( r_{m},x_{h,m},x_{l,m}\right) =\Omega _{i} \\ 
=0 & \text{if} & U_{i,m}\left( r_{m},x_{h,m},x_{l,m}\right) <\Omega _{i}%
\end{array}%
\right. .  \label{Buyers indifferent across mechanisms}
\end{equation}%
In words, the tuple $\left( r_{m},x_{h,m},x_{l,m}\right) $ must generate an
expected utility of at least $\Omega _{h}$ for high type customers, else
they will stay away ($x_{h,m}=0$) and at least $\Omega _{l}$ for low type
customers, else they will stay away ($x_{l,m}=0).$When competing for buyers,
each seller chooses a rule $m=s,ap$ and a price $r_{m}\in \left[ 0,1\right] $
but realizes that expected demands $x_{h,m}$ and $x_{l,m}$ are determined
via (\ref{Buyers indifferent across mechanisms}).

\begin{lemma}
\label{Lemma: no 1st or 2nd}Fix some arrival rates $x_{h}$, $x_{l}$ and a
reserve price $r.$ The all-pay format generates strictly lower payoffs for
buyers than first-price or second-price formats. Equivalently, it generates
strictly higher profits for the seller.
\end{lemma}

\textbf{Proof of Lemma \ref{Lemma: no 1st or 2nd}}. Let $x=x_{h}+x_{l}$ and $%
\theta =x_{l}/x$, therefore $x_{l}=\theta x$ and $x_{h}=\left( 1-\theta
\right) x.$ Recall that under standard auctions%
\begin{equation*}
u_{h}\left( n\right) =\theta ^{n-1}\left( 1-b\right) \ \ \ \ \ \text{and}\ \
\ \ u_{l}\left( n\right) =\frac{\theta ^{n-1}}{n}\left( 1-b\right) \text{
for }n\geq 2,
\end{equation*}%
whereas with all pay auctions%
\begin{equation*}
u_{h}\left( n\right) =\max \left( \theta ^{n-1}-b,0\right) \text{ \ and \ \ }%
u_{l}\left( n\right) =\max \left( \theta ^{n-1}/n-b,0\right) \text{ for }%
n\geq 2.
\end{equation*}%
Further recall that%
\begin{equation*}
U_{i}=z_{0}\left( x\right) \left( 1-r\right) +\sum_{n=1}^{\infty
}z_{n}\left( x\right) u_{i}\left( n+1\right) .
\end{equation*}%
Noting that $r$ and $z_{n}\left( x\right) $s are the same across all rules,
the claim that the all-pay format generates strictly lower payoffs for
buyers than first-price or second-price formats entails showing that%
\begin{equation*}
\theta ^{n-1}\left( 1-b\right) >^{?}\max \left( \theta ^{n-1}-b,0\right) 
\text{ \ and \ \ }\frac{\theta ^{n-1}}{n}\left( 1-b\right) >^{?}\max \left( 
\frac{\theta ^{n-1}}{n}-b,0\right) .
\end{equation*}%
It is straightforward to check that both inequalities hold. The claim that
sellers earn more under all pay auction follows from the fact that%
\begin{equation*}
\Pi =1-z_{0}\left( x\right) -x_{l}U_{l}-x_{h}U_{h}.
\end{equation*}%
Since $x$s are controlled for, the fact that $U$s are smaller with all pay
auctions implies that the $\Pi $ is higher. $\blacksquare $

\begin{proposition}
\label{prop-no 1st2ndeq'm}. A first or second price auction equilibrium (in
which all sellers adopt these rules)\ fails to exist. In such a scenario a
seller can unilaterally switch to all pay auctions and earn more.
\end{proposition}

\textbf{Proof of Proposition \ref{prop-no 1st2ndeq'm}}. Consider an
equilibrium in which sellers adopt a standard\ (first or second price)
format. Symmetry in buyers' visiting strategies implies that each seller
receives $\lambda _{l}$ low types, $\lambda _{h}$ high types, and it total $%
\lambda _{l}+\lambda _{h}=\lambda $ customers. Recall that the percentage of
low types is $\sigma .$ The expected payoffs and profits are 
\begin{align*}
U_{h,s}& =z_{0}\left( \lambda \right) \left( 1-r_{s}\right) +z_{0}\left(
\lambda _{h}\right) \left( 1-z_{0}\left( \lambda _{l}\right) \right) \left(
1-b\right) \text{ and } \\
U_{l,s}& =z_{0}\left( \lambda \right) \left( 1-r_{s}\right) +z_{0}\left(
\lambda _{h}\right) \frac{1-z_{0}\left( \lambda _{l}\right) -z_{1}\left(
\lambda _{l}\right) }{\lambda _{l}}\left( 1-b\right) .
\end{align*}%
and therefore%
\begin{equation*}
\Pi _{s}=1-z_{0}\left( \lambda \right) -\lambda \left( 1-\sigma \right)
U_{h,s}-\sigma \lambda U_{l,s}.
\end{equation*}%
Now consider a seller who switches to all pay auctions. We will show that
this seller can keep providing high types with $U_{h,s}$ and low types with $%
U_{l,s}$ yet can earn more.

Once the seller makes such a switch, there are three key parameters:\ the
reserve price $r_{a}$, the total demand, $x,$ and the composition of demand, 
$\theta .$ The fact that these parameters are potentially different than $%
r_{s},$ $\lambda $ and $\sigma $ makes the comparison difficult. To get
around this issue, we will control for $\lambda $, and show that there there
exists a reserve price $\hat{r}<r_{s}$ and $\theta >\sigma $ ensuring that $%
U_{l,a}=U_{l,s}$ and $U_{h,a}=U_{h,s},$ i.e. the seller attracts $\lambda $
customers in total\ (albeit with a different composition $\theta $ and
reserve $r)$ and provides low types the same payoff $U_{l,s}$ and high types
with the same payoff$.$

Per Lemma \ref{Lemma: no 1st or 2nd} we have%
\begin{equation*}
U_{h,a}\left( r,\lambda ,\theta \right) <U_{h,s}\left( r_{s},\lambda ,\sigma
\right) \text{ \ \ and \ \ }U_{l,a}\left( r,\lambda ,\theta \right)
<U_{l,s}\left( r_{s},\lambda ,\sigma \right) \text{ when }r=r_{s}\text{ and }%
\theta =\sigma .
\end{equation*}%
Note that both $U_{h,a}$ and $U_{l,a}$ fall in $r.$ In addition under all
pay auctions%
\begin{equation*}
u_{h}=\max \left( \theta ^{n-1}-b,0\right) \text{ and }u_{l}=\max \left(
\theta ^{n-1}/n-b,0\right)
\end{equation*}%
which means that both $U_{h,a}$ and $U_{l,a}$ rise in $\theta .$ It follows
that there exists some $\hat{\theta}>\sigma $ and $\hat{r}<r_{s}$ satisfying 
$U_{l,a}=U_{l,s}$ \ \ and \ \ $U_{h,a}=U_{h,s}$ Now, the seller's payoff is%
\begin{equation*}
\Pi _{a}=1-z_{0}\left( \lambda \right) -\lambda (1-\hat{\theta})U_{h,s}-\hat{%
\theta}\lambda U_{l,s}
\end{equation*}%
Since $U_{h,s}>U_{l,s}$ and $\hat{\theta}>\sigma ,$ we have $\Pi _{a}>\Pi
_{s},$ i.e. deviation is profitable. $\blacksquare $

In words, the seller attracts the same total demand $\lambda ,$ yet the
buyer mix shifts with $\hat{\theta}>\sigma $ implying an increase in
low-type buyers and a decrease in high-type ones. High types are paid higher
utility than low types, so this shift benefits the seller, enabling him to
generate more earnings.

\begin{proposition}
\label{prop - main}If $b>\sigma $ then in the unique symmetric equilibrium
all sellers select the all-pay auction format.
\end{proposition}

The proof outlines the all-pay equilibrium: Sellers set $r=0$, and all
buyers, both low and high types, achieve the identical expected utility $%
z_{0}\left( \lambda \right) $. Remarkably, this is the same outcome in a
model with homogeneous buyers\ (with no budget constraints). Due to the
structure of the all-pay auction, participants place bids of small amounts,
and therefore, those with lower financial capabilities still engage and, on
expected terms, earn as much as they would without any budget constraints.

\section{Conclusion}

In an all-pay auction, every participant must pay their bids, irrespective
of winning or losing. This setup encourages lower bids, opening avenues for
individuals with limited budgets to participate. Lower bids don't
necessarily translate to reduced revenues since the seller gathers all bids,
not just the winning one. In fact, we demonstrate that in a similar context,
the all-pay format outperforms both first and second-price formats in terms
of revenue. This difference gives the all-pay rule a competitive edge over
other formats in a model where stores compete for customers.

\bibliographystyle{apalike}
\bibliography{acompat,References_Allpay}
\bigskip\ 

\pagebreak

\section{Appendix}

\textbf{Proof of the Proposition \ref{Proposition-AuctionOutcome-ALLPAY}. }%
We focus on symmetric mixed strategies where bidders of the same type pick
the same cumulative distribution function (cdf) $G_{i}\left( p\right) :\left[
\underline{s}_{i},\bar{s}_{i}\right] \rightarrow \left[ 0,1\right] $. A
point $p$ is an `increasing point' of $G_{i}$ if $G_{i}$ is not constant in
an $\varepsilon $ neighborhood of $p_{i}$ i.e. if for each $\varepsilon >0$
the probability of having a value in $(p-\varepsilon ,p+\varepsilon )$ is
positive. Conversely, $p$ is a `constant point' if $G_{i}$ is constant in an 
$\varepsilon $ neighborhood of $p$. Note that if there is an atom at $p$
then, by definition, $p$ is an increasing point. If the pair $\left(
G_{h},G_{l}\right) $ corresponds to an equilibrium, then type $i$ buyers
earn their equilibrium payoff $u_{i}$ at each increasing point $p$ of $%
G_{i}. $ Similarly they earn an expected payoff that is less than or equal
to $u_{i} $ at each constant point $p$ of $G_{i}$ %
\citep{hillman_dissipation_1987,baye_all-pay_1996}. Said differently, in a
mixed strategy equilibrium players must be indifferent to all increasing
points and they must weakly prefer increasing points over constant points in
the support of their bidding distribution. These claims are immediate from
the definition of a mixed strategy equilibrium; \cite%
{hillman_dissipation_1987} provide a more formal discussion (see Proposition
2 therein).

The equilibrium cdf $G_{h}$ cannot have a mass point anywhere on its support 
$\left[ \underline{s}_{h},\bar{s}_{h}\right] .$ A mass point means tying
with other bidders in which case the surplus is divided among the tying
bidders via random rationing. A financially unconstrained bidder can always
beat the tie and improve his payoff by placing a bid that is slightly above
the mass point, which is inconsistent with $G_{h}$ being an equilibrium
distribution. The argument applies to the entire support including the upper
bound $\bar{s}_{h}$:\ If $\bar{s}_{h}<1$ then there is room to beat a
potential tie at $\bar{s}_{h}$. If $\bar{s}_{h}=1$ then a mass at $1$ would
result in a negative payoff.\ In either case there cannot be a mass at $\bar{%
s}_{h}$. To sum up $G_{h}$ is continuos on its support with no jumps (it
does not have flat spots either, but more on this below).

The equilibrium cdf $G_{l}$ cannot have an atom anywhere below $b$ for the
same reason above, however it may have an atom at $b$. There are three
scenarios for $G_{l}:$

L1 --- The entire mass is at $b.$

L2 --- Partial mass at $b$ with a continuos tail below $b.$

L3 --- No mass at $b,$ so the entire cdf is continuos.

Before we move on, the following expression will be useful. Assuming that $p$
is not a mass point$,$ the expected payoff associated with bidding $p$ is
given by 
\begin{eqnarray}
EU\left( p\right) &=&\sum_{i=0}^{n-1}\binom{n-1}{i}\left[ \theta G_{l}\left(
p\right) \right] ^{i}\left[ \left( 1-\theta \right) G_{h}\left( p\right) %
\right] ^{n-1-i}-p  \notag \\
&=&\left[ \theta G_{l}\left( p\right) +\left( 1-\theta \right) G_{h}\left(
p\right) \right] ^{n-1}-p.  \label{EU(p) - ALLPAY}
\end{eqnarray}%
This is true for both types of buyers. Indeed buyers have identical
valuations for the item, so both types earn the same expected payoff $%
EU\left( p\right) ,$ assuming, of course, $p\leq b$. If $p>b$ then low types
are sure to be outbid; the expected payoff for high types can be obtained by
substituting $G_{l}\left( p\right) =1$ into above.

\begin{itemize}
\item \textbf{Region 1}:\ Suppose that $b<\frac{\theta ^{n-1}}{n}$
\end{itemize}

1A. In this parameter region high types are guaranteed to receive a payoff $%
\theta ^{n-1}-b$ whereas low types are guaranteed to receive $\frac{\theta
^{n-1}}{n}-b$, both of which are positive because $b<\frac{\theta ^{n-1}}{n}%
. $ Low types cannot bid more than their budget $b,$ so if a high type buyer
bids slightly more than $b,$ then even if he looses against all other high
types, he can still win the item with probability $\theta ^{n-1}$ (everyone
else is a low type) and would obtain a payoff $\theta ^{n-1}-b.$ Similarly
if a low type bids $b,$ then in the worst case scenario he looses against
all high types and ties with every other low type, so his payoff is at least 
$\frac{\theta ^{n-1}}{n}-b.$ It follows that $u_{h}\geq u_{l}>0.$

1B. We rule out scenarios L2 and L3, which leaves L1 as the only possible
scenario for $G_{l}.$ In scenarios L2 and L3 the cdf $G_{l}$ is assumed to
be continuos over some interval $\left[ \underline{s}_{l},\bar{s}_{l}\right] 
$ where $0\leq \underline{s}_{l}<\bar{s}_{l}\leq b$. Recall that $G_{h}$ is
also continuos over $\left[ \underline{s}_{h},\bar{s}_{h}\right] ,$ so there
are three possibilities:\ either \underline{$s$}$_{h}>$\underline{$s$}$_{l}$
or \underline{$s$}$_{l}>$\underline{$s$}$_{h}$ or \underline{$s$}$_{l}=$%
\underline{$s$}$_{h}.$ Consider the first one, i.e. suppose \underline{$s$}$%
_{h}>\underline{s}_{l}.$ This implies $u_{l}=-\underline{s}_{l}.$ To see
why, note that \underline{$s$}$_{l}$ is an increasing point of $G_{l}$ and
in a mixed strategy equilibrium any increasing point, including \underline{$%
s $}$_{l},\,$must yield the equilibrium payoff to the bidder. If the buyer
bids $p=\underline{s}_{l}$ then he is sure to lose the item as everyone else
is sure to bid more than\ \underline{$s$}$_{l}$ (recall that \underline{$s$}$%
_{l}<$\underline{$s$}$_{h}$ ). His payoff, therefore, is $-\underline{s}_{l}$
$\leq 0$ because in the all-pay auction all participants forfeit their bids
(if the lower bound $\underline{s}_{l}$ is set to be zero then the resulting
payoff is zero, but if $\underline{s}_{l}$ is positive then the payoff is
negative). This, of course, contradicts the fact that $u_{l}>0.$ Now suppose 
\underline{$s$}$_{l}>$\underline{$s$}$_{h}.$ This implies $u_{h}=-\bar{s}%
_{h}\leq 0,$ which contradicts $u_{h}>0$. Finally if \underline{$s$}$_{l}=$%
\underline{$s$}$_{h}$ then $u_{h}=u_{l}=-\underline{s}_{l}\leq 0,$ which
again is a contradiction. In words, if both cdfs have continuos bits then
the one with the lower bound on the far left is bound to yield at most a
zero payoff. It follows that $G_{l}$ cannot have a continuos part over some
interval below $b;$ the entire mass must be at point $b.$

1C. Given that the entire mass of $G_{l}$ is placed at $b$ the lower bound
of $G_{h}$ cannot be below $b,$ i.e. we must have \underline{$s$}$_{h}\geq
b. $ Indeed if \underline{$s$}$_{h}<b$ then $u_{h}=-\underline{s}_{h}\leq 0,$
a contradiction. Since all low types bid $b$ and high types are sure to
outbid the low types\ (\underline{$s$}$_{h}\geq b)$ the equilibrium payoff
of a low type equals to $u_{l}=\frac{\theta ^{n-1}}{n}-b.$

1D. As discussed earlier, $G_{h}$ cannot have an atom, i.e. there are no
jumps. We now argue that it cannot have intermittent flat spots either. By
contradiction, suppose $G_{h}$ is constant at some interval $\left(
a_{1},a_{2}\right) \subset \lbrack $\underline{$s$}$_{h},\bar{s}_{h}].$ Both 
$a_{1}$ and $a_{2}$ are increasing points of the distribution function $%
G_{h} $ hence they both must deliver the same payoff $u_{h}.$ Since $G_{h}$
is flat in this interval the probability of winning the auction is the same
at both points (notice also $G_{l}=1$ at both points). This, however, means
that the player gets a lower payoff at $a_{2}$ than $a_{1}$ since $%
a_{2}>a_{1};$ so, he cannot be indifferent; a contradiction.

1E. We can now characterize $G_{h}.$ We established that $G_{h}$ is
monotonically increasing on its support $[$\underline{$s$}$_{h},\bar{s}%
_{h}]. $ The expected payoff associated with bidding any $p\in \lbrack $%
\underline{$s$}$_{h},\bar{s}_{h}]$ is given by%
\begin{equation*}
EU\left( p\right) =\left[ \theta +\left( 1-\theta \right) G_{h}\left(
p\right) \right] ^{n-1}-p,
\end{equation*}%
which is obtained by substituting $G_{l}\left( p\right) =1$ into (\ref{EU(p)
- ALLPAY}) (since \underline{$s$}$_{h}\geq b$ we have $G_{l}\left( p\right)
=1$ for all $p\geq $\underline{$s$}$_{h}$). High types must earn their
equilibrium payoff $u_{h}$ at any increasing point of $G_{h}$. Since $G_{h}$
is monotonically increasing we must have 
\begin{equation*}
EU\left( p\right) =u_{h}\text{ for all }p\in \lbrack \underline{s}_{h},\bar{s%
}_{h}].
\end{equation*}%
Substituting for $EU\left( p\right) $ and solving for $G_{h}$ we have 
\begin{equation*}
G_{h}\left( p\right) =\frac{\left( p+u_{h}\right) ^{\frac{1}{n-1}}-\theta }{%
1-\theta }.
\end{equation*}%
We know $G_{h}\left( \underline{s}_{h}\right) =0$ and $G_{h}\left( \bar{s}%
_{h}\right) =1,$ hence $u_{h}=\theta ^{n-1}-\underline{s}_{h}$\ and $u_{h}=1-%
\bar{s}_{h}.$ Recall that $u_{h}\geq \theta ^{n-1}-b\ $and that \underline{$%
s $}$_{h}\geq b.$ This means \underline{$s$}$_{h}=b$ and $u_{h}=\theta
^{n-1}-b $ and therefore $\bar{s}_{h}=1-\theta ^{n-1}+b.$ Substituting for $%
u_{h}$ yields the expression of $G_{h}$ in the body of the proposition.

\begin{itemize}
\item \textbf{Region 2}:\ Suppose that $\frac{\theta ^{n-1}}{n}<b<\theta
^{n-1}$
\end{itemize}

2A. Per the discussion in 1A, high types are still guaranteed to receive a
positive payoff, $\theta ^{n-1}-b;$ however, low types are no longer
guaranteed to receive a positive payoff i.e$.$ $u_{h}>0,$ $u_{l}\geq 0.$

2B. There are three scenarios for $G_{l}$ which are outlined in L1, L2 and
L3. Suppose L1 is valid, i.e. low types bid $b$ for sure. Then \underline{$s$%
}$_{h}\geq b$ and the equilibrium payoff of a low type equals to $u_{l}=%
\frac{\theta ^{n-1}}{n}-b$ (see 1C). This, however, is negative because $%
\frac{\theta ^{n-1}}{n}<b;$ a contradiction. Since L1 is ruled out, we have
either L2 or L3. In either case $G_{l}$ is continuos over some interval $%
\left[ \underline{s}_{l},\bar{s}_{l}\right] $ where $0\leq \underline{s}_{l}<%
\bar{s}_{l}\leq b.$ Recall that $G_{h}$ is continuos on its support $\left[ 
\underline{s}_{h},\bar{s}_{h}\right] $, so there are three possibilities:\
either \underline{$s$}$_{h}>$\underline{$s$}$_{l}$ or \underline{$s$}$_{l}>$%
\underline{$s$}$_{h}$ or \underline{$s$}$_{l}=$\underline{$s$}$_{h}.$ Per
the discussion in 1B, we cannot have \underline{$s$}$_{l}>$\underline{$s$}$%
_{h}$ or \underline{$s$}$_{l}=$\underline{$s$}$_{h}$ because in either case $%
u_{h}\leq 0,$ a contradiction. So we must have \underline{$s$}$_{h}>$%
\underline{$s$}$_{l}.$ This implies $u_{l}=-$\underline{$s$}$_{l}\leq 0.$
Clearly if \underline{$s$}$_{l}>0$ then $u_{l}$ is negative; hence we must
have \underline{$s$}$_{l}=0$ and therefore $u_{l}=0.$

2C. We will show that \underline{$s$}$_{h}\geq \bar{s}_{l},$ i.e. the
supports of $G_{h}$ and $G_{l}$ cannot overlap. Suppose they do, i.e.
suppose that \underline{$s$}$_{h}<\bar{s}_{l}$ so that $G_{h}\left( \bar{s}%
_{l}\right) >0.$ Pick some point $p\in \left[ \underline{s}_{h},\bar{s}_{l}%
\right] $ at which $G_{h}$ is increasing (since $G_{h}\left( \bar{s}%
_{l}\right) >0$ such a point must exist). Recall that the expected payoff
associated with bidding $p$ is equal to $EU\left( p\right) $, which is given
by (\ref{EU(p) - ALLPAY}), and it is the same for both types of buyers.
Since $p$ is an increasing point of $G_{h}$ we have $EU\left( p\right)
=u_{h}.$ Now $p$ is either an increasing point of $G_{l}$ or it is a
constant point of $G_{l}.$ The first case implies $EU\left( p\right) =u_{l}$
whereas the second one implies $EU\left( p\right) \leq u_{l}.$ In either
case we have a contradiction since $u_{h}>0$ and $u_{l}=0.$ It follows that
the supports of $G_{h}$ and $G_{l}$ cannot overlap, so we must have 
\underline{$s$}$_{h}\geq \bar{s}_{l}.$

2D. Now we will rule out scenario L3. Again, by contradiction, suppose L3 is
valid, i.e. $G_{l}$ is atomless and does not jump anywhere on its support $%
\left[ 0,\bar{s}_{l}\right] .$ In 2C we proved that \underline{$s$}$_{h}\geq 
\bar{s}_{l},$ hence $G_{h}\left( p\right) =0$ for all $p\in \left[ 0,\bar{s}%
_{l}\right] .$ Furthermore, per the discussion in 1D, $G_{l}$ cannot have
intermittent flat spots either; hence $G_{l}$ is monotonically increasing on
its support$.$ The expected payoff associated with bidding any $p\in \lbrack
0,\bar{s}_{l}]$ is given by%
\begin{equation*}
EU\left( p\right) =\left[ \theta G_{l}\left( p\right) \right] ^{n-1}-p,
\end{equation*}%
which is obtained by substituting $G_{h}\left( p\right) =0$ into (\ref{EU(p)
- ALLPAY}). Low types must earn their equilibrium payoff $u_{l}=0$ at any
increasing point of $G_{l}$. Since $G_{l}$ is monotonically increasing we
must have $EU\left( p\right) =u_{h}$ for all $p\in \lbrack \underline{s}_{h},%
\bar{s}_{h}].$ Substituting for $EU\left( p\right) $ and solving for $G_{l}$
we have 
\begin{equation}
G_{l}\left( p\right) =\frac{p^{\frac{1}{n-1}}}{\theta }.
\label{GL - Appendix}
\end{equation}%
Note that $G_{l}\left( \bar{s}_{l}\right) $ must be equal to 1; however this
is impossible because $\bar{s}_{l}\leq b$ and $G_{l}\left( b\right) =\frac{%
b^{\frac{1}{n-1}}}{\theta }<1$ since $b<\theta ^{n-1}$. Without a mass point
at $b$, the function $G_{l}$ cannot be a valid cdf.

2E. Since L3 is ruled out, the only possible scenario is L2 where $G_{l}$
has some partial mass $\mu $ at point $b$ while the remaining mass is spread
over some interval starting at the lower bound \underline{$s$}$_{l}=0.$ Per
the discussion in 2D, at any point of increase in the region where $G_{l}$
is atomless low types must earn $u_{l}=0,$ which implies $G_{l}\left(
p\right) =\frac{p^{\frac{1}{n-1}}}{\theta }.$ Buyers should get the same
payoff $u_{l}=0$ at the mass point $b$ as well$.$ The expected payoff
associated with bidding $b$ is given by%
\begin{equation*}
\theta ^{n-1}\sum_{i=0}^{n-1}\binom{n-1}{i}\frac{\mu ^{i}}{i+1}\left( 1-\mu
\right) ^{n-i-1}-b=\frac{\theta ^{n-1}\left[ 1-\left( 1-\mu \right) ^{n}%
\right] }{\mu n}-b.
\end{equation*}%
For this to be equal to zero, $\mu $ must solve%
\begin{equation*}
\frac{bn}{\theta ^{n-1}}-\frac{1-\left( 1-\mu \right) ^{n}}{\mu }=0
\end{equation*}%
It is straightforward to show that so as long as $\frac{\theta ^{n-1}}{n}$ $%
<b<\theta ^{n-1}b$ there exists a unique $\mu \in \left( 0,1\right) $
satisfying above. The upper bound of the atomless portion of $G_{l},$ call
it $\bar{p}$, satisfies $G_{l}\left( \bar{p}\right) =1-\mu $ hence 
\begin{equation*}
\bar{p}=\theta ^{n-1}\left( 1-\mu \right) ^{n-1}.
\end{equation*}%
An argument similar to the one in 1D reveals that $G_{l}$ cannot have flat
spots in the region $\left( 0,\bar{p}\right) .$ So $G_{l}$ monotonically
rises in $\left[ 0,\bar{p}\right] $, has mass point at $b$ and it is flat in
between.

2F. We now characterize $G_{h}.$ We know \underline{$s$}$_{h}\geq \bar{s}%
_{l}=b,$ so for any $p\geq $\underline{$s$}$_{h}$ we have $G_{l}\left(
p\right) =1.$ At any point of increase in the support of $G_{h}$ we must
have $EU\left( p\right) =u_{h}$ where 
\begin{equation*}
EU\left( p\right) =\left[ \theta +\left( 1-\theta \right) G_{h}\left(
p\right) \right] ^{n-1}-p,
\end{equation*}%
which is obtained by substituting $G_{l}\left( p\right) =1$ into\ (\ref%
{EU(p) - ALLPAY}). It follows that 
\begin{equation*}
G_{h}\left( p\right) =\frac{\left( p+u_{h}\right) ^{\frac{1}{n-1}}-\theta }{%
1-\theta }.
\end{equation*}%
Since $G_{h}\left( \underline{s}_{h}\right) =0$ we have \underline{$s$}$%
_{h}=\theta ^{n-1}-u_{h}.$ Since $u_{h}\geq \theta ^{n-1}-b$ and \underline{$%
s$}$_{h}\geq b$ we have \underline{$s$}$_{h}=b$ and $u_{h}=\theta ^{n-1}-b$
and therefore $\bar{s}_{h}=1-\theta ^{n-1}+b.$ Per the discussion in 1D, $%
G_{h}$ cannot have intermittent flat spots anywhere in its support.

\begin{itemize}
\item \textbf{Region 3}:\ Suppose $\theta ^{n-1}\leq b.$
\end{itemize}

3A. Since $\theta ^{n-1}\leq b$, per 1A, buyers are no longer guaranteed to
receive a positive payoff. In regions 1 and 2 at least one of the
equilibrium payoffs was positive, and this feature played a key role in
establishing the uniqueness of the equilibrium. Without this information
proving uniqueness becomes a challenge; so, instead of attempting to
characterize the equilibrium, we focus on expected payoffs and prove that in
any symmetric equilibrium the seller extracts the entire surplus, i.e. $%
u_{l}=u_{h}=0$. We then verify that the strategy profile in (\ref%
{GL-GH-ZeroPayoffs}) constitutes such an equilibrium.

3B. Consider a symmetric mixed strategy equilibrium with cdfs $G_{h}$, $%
G_{l} $ and associated equilibrium payoffs $u_{h},$ $u_{l}$. WLOG let $%
u_{h}\geq u_{l},$ so there are three possibilities:\ (i) $u_{h}\geq u_{l}>0$
or (ii)$\ u_{h}>u_{l}=0$ or (iii) $u_{h}=u_{l}=0.$ We will rule out (i) and
(ii). To start, suppose $u_{h}\geq u_{l}>0$. Recall that $G_{h}$ is atomless
whereas there are three scenarios for $G_{l}$. Per 1B, if both cdfs have
continuos bits then the one with the lower bound on the far left will yield
at most a zero payoff. So, if $u_{l}$ is positive then $G_{l}$ cannot have a
continuos part over some interval below $b;$ the entire mass must be at
point $b$. This scenario is analyzed in 1C and the equilibrium payoff of a
low type equals to $u_{l}=\frac{\theta ^{n-1}}{n}-b,$ which is negative
since $\theta ^{n-1}\leq b;$ a contradiction.

3C. Suppose $u_{h}>u_{l}=0$. The cdf $G_{l}$ cannot have the entire mass at $%
b$ (3B). So either it has a partial mass at $b$ or it is atomless
everywhere. Suppose it has a partial mass at $b.$ Then, per 2C, \underline{$%
s $}$_{h}\geq \bar{s}_{l}=b$ (cdfs cannot overlap). For a high type the
expected payoff associated with bidding \underline{$s$}$_{h}$ equals to $EU(%
\underline{s}_{h})=\theta ^{n-1}-b$, which is less than or equal to zero
since $\theta ^{n-1}\leq b.$ The lower bound \underline{$s$}$_{h}$ is an
increasing point of $G_{h},$ so we must have $EU(\underline{s}_{h})=u_{h}.$
This, however, is a contradiction because $u_{h}>0$ but $EU(\underline{s}%
_{h})\leq 0.$ The final scenario for $G_{l}$ is where it is atomless
everywhere; so, suppose this is the case. Per 2B, 2C and 2D, $G_{l}$ is
given by (\ref{GL - Appendix}) and it must be monotonically increasing on
its support $\left[ 0,\bar{s}_{l}\right] $ with no flat spots. Solving $%
G_{l}\left( \bar{s}_{l}\right) =1$ yields $\bar{s}_{l}=\theta ^{n-1}.$
Recall that \underline{$s$}$_{h}\geq \bar{s}_{l}$, so \underline{$s$}$%
_{h}\geq \theta ^{n-1}$. It follows that $EU(\underline{s}_{h})\leq 0,$
which, again is a contradiction since we must have $EU(\underline{s}%
_{h})=u_{h}>0.$

3D. Arguments in 3B and 3C imply that we must have $u_{l}=u_{h}=0.$ What
remains to be done is to characterize equilibrium strategies $G_{l}$ and $%
G_{h}$ delivering these payoffs. For the purpose of the paper the fact that
any symmetric equilibrium yields zero payoffs is sufficient, so we refrain
from attempting to characterize all possible combinations of $G_{l}$ and $%
G_{h};$ instead we will verify that the specific forms of $G_{l}$ and $%
G_{h}, $ given by\ (\ref{GL-GH-ZeroPayoffs}), indeed correspond to an
equilibrium and they yield zero payoffs. So suppose all players adopt the
cdfs in (\ref{GL-GH-ZeroPayoffs}) and consider a potential deviation by a
low type buyer who picks a different cdf, say, $\tilde{G}:\left[ 0,b\right]
\rightarrow \left[ 0,1\right] .$ His expected payoff $\tilde{u}$ is given by%
\begin{equation*}
\tilde{u}=\int_{0}^{b}\{[\theta G_{l}\left( p\right) +\left( 1-\theta
\right) G_{h}\left( p\right) ]^{n-1}-p\}d\tilde{G}\left( p\right) .
\end{equation*}%
Observe that $G_{l}\left( p\right) =1$ for $p\geq \theta ^{n-1}$ and $%
G_{h}\left( p\right) =0$ for $p\leq \theta ^{n-1};$ therefore%
\begin{equation*}
\tilde{u}=\int_{0}^{\theta ^{n-1}}\{\theta ^{n-1}G_{l}^{n-1}\left( p\right)
-p\}d\tilde{G}\left( p\right) +\int_{\theta ^{n-1}}^{b}\{[\theta +\left(
1-\theta \right) G_{h}\left( p\right) ]^{n-1}-p\}d\tilde{G}\left( p\right) .
\end{equation*}%
After substituting for $G_{l}\left( p\right) $ and $G_{h}\left( p\right) $,
which are given by (\ref{GL-GH-ZeroPayoffs}), the expressions inside the
curly brackets vanish, thus $\tilde{u}=0$ irrespective of $\tilde{G},$ i.e.
if everyone else sticks to (\ref{GL-GH-ZeroPayoffs}) then a low type cannot
earn anything but zero irrespective of the cdf he picks. A similar argument
applies to high types as well. This completes the proof. $\blacksquare $

\bigskip

\noindent \textbf{Proof of Lemma \ref{Lemma - Pi and U}.} Start with
standard auctions (denoted with subscript $s)$. If a buyer is alone at an
auction store then he obtains the item by paying the reserve price i.e. $%
u_{h,s}\left( 1\right) =u_{l,s}\left( 1\right) =1-r_{s}.$ If $n\geq 2$ then
we know that under both auction formats%
\begin{equation*}
u_{h,s}\left( n\right) =\theta ^{n-1}\left( 1-b\right) \ \text{and}\
u_{l,s}\left( n\right) =\frac{\theta ^{n-1}}{n}\left( 1-b\right) ,\text{
where }\theta =\frac{x_{l,s}}{x_{h,s}+x_{l,s}}.
\end{equation*}%
Substituting these expressions into (\ref{Uim - RAw}) yields 
\begin{eqnarray*}
U_{h,s} &=&z_{0}\left( x_{h,s}+x_{l,s}\right) \left( 1-r_{s}\right)
+\sum_{n=1}^{\infty }z_{n}\left( x_{h,s}+x_{l,s}\right) \theta ^{n}\left(
1-b\right) \text{ and } \\
U_{l,s} &=&z_{0}\left( x_{h,s}+x_{l,s}\right) \left( 1-r_{s}\right)
+\sum_{n=1}^{\infty }z_{n}\left( x_{h,s}+x_{l,s}\right) \frac{\theta ^{n}}{%
n+1}\left( 1-b\right) .
\end{eqnarray*}%
After substituting for $\theta $ and re-arranging we get%
\begin{align}
U_{h,s}& =z_{0}\left( x_{h,s}+x_{l,s}\right) \left( 1-r_{s}\right)
+z_{0}\left( x_{h,s}\right) \left( 1-z_{0}\left( x_{l,s}\right) \right)
\left( 1-b\right) \text{ and }  \label{acik artirma kalantor indiffrence} \\
U_{l,s}& =z_{0}\left( x_{h,s}+x_{l,s}\right) \left( 1-r_{s}\right)
+z_{0}\left( x_{h,s}\right) \frac{1-z_{0}\left( x_{l,s}\right) -z_{1}\left(
x_{l,s}\right) }{x_{l,s}}\left( 1-b\right) .
\label{acik artirma culsuz indiffrence}
\end{align}

With all pay auctions, assuming $b>\theta ,$ we have%
\begin{equation*}
u_{h,ap}\left( 1\right) =u_{l,ap}\left( 1\right) =1-r_{ap}\text{ \ and \ }%
u_{h,ap}\left( n\right) =u_{l,ap}\left( n\right) =0\text{ for }n\geq 2.
\end{equation*}%
Substituting this into\ (\ref{Uim - RAw}) yields 
\begin{equation}
U_{h,ap}=U_{l,ap}=z_{0}(x_{h,ap}+x_{l,ap})(1-r_{ap}).  \label{Ua}
\end{equation}%
Now we can link these to the profit functions. Start with standard auctions.
If a single customer is present then the reserve price is charged, i.e. $\pi
_{s}\left( 1\right) =r_{s}.$ If $n\geq 2$ then both auction formats yield
the same $\pi _{s}\left( n\right) $, given by (\ref{pi(n) - 1st 2nd}); hence 
\begin{equation*}
\Pi _{s}=z_{1}\left( x_{h,s}+x_{l,s}\right) r_{s}+\sum_{n=2}^{\infty
}z_{n}\left( x_{h,s}+x_{l,s}\right) \{[\theta ^{n}+n\theta ^{n-1}\left(
1-\theta \right) ]b+1-\theta ^{n}-n\theta ^{n-1}\left( 1-\theta \right) \}.
\end{equation*}%
After substituting for $z_{n}$ and $\theta $ and simplifying we have%
\begin{equation}
\begin{array}{cl}
\Pi _{a}= & z_{1}\left( x_{h,s}+x_{l,s}\right) r_{s}+1-z_{0}\left(
x_{h,s}\right) -z_{1}\left( x_{h,s}\right) \bigskip \\ 
& +b\{z_{0}\left( x_{h,s}\right) +z_{1}\left( x_{h,s}\right) -z_{0}\left(
x_{h,s}+x_{l,s}\right) -z_{1}\left( x_{h,s}+x_{l,s}\right) \}.%
\end{array}
\label{Profit Auctions}
\end{equation}%
Note that $z_{1}\left( x\right) =xz_{0}\left( x\right) $ and $z_{0}\left(
x+y\right) =z_{0}\left( x\right) z_{0}\left( y\right) $. It follows that 
\begin{eqnarray*}
x_{h,s}U_{h,s}+x_{l,s}U_{l,s} &=&z_{0}\left( x_{h,s}\right) +z_{1}\left(
x_{h,s}\right) -z_{0}\left( x_{h,s}+x_{l,s}\right) -z_{1}\left(
x_{h,s}+x_{l,s}\right) r_{s}+ \\
&&-b\left\{ z_{0}\left( x_{h,s}\right) +z_{1}\left( x_{h,s}\right)
-z_{0}\left( x_{h,s}+x_{l,s}\right) -z_{1}\left( x_{h,s}+x_{l,s}\right)
\right\} ,
\end{eqnarray*}%
where $U_{h,s}$ and $U_{l,s}$ are given by (\ref{acik artirma kalantor
indiffrence})\ and (\ref{acik artirma culsuz indiffrence}). The expected
profit $\Pi _{a}$ is given by (\ref{Profit Auctions}). A term by term
comparison reveals that%
\begin{equation*}
\Pi _{a}=1-z_{0}\left( x_{h,s}+x_{l,s}\right) -x_{h,s}U_{h,s}-x_{l,s}U_{l,s},
\end{equation*}%
confirming the validity of the relationship under auctions.

As for all pay auctions, suppose $b>\theta .$ Then we will prove that the
seller's expected earning is $\pi \left( n\right) =1.$ First we calculate
the expected bids from low and high types. For low types this is given by%
\begin{equation*}
\int_{0}^{\theta ^{n-1}}pdG_{l}\left( p\right) \text{ \ \ \ \ }=\text{ \ \ \
\ }\left. pG_{l}\left( p\right) \right\vert _{0}^{\theta
^{n-1}}-\int_{0}^{\theta ^{n-1}}G_{l}\left( p\right) dp\text{ \ \ \ \ }=%
\text{ \ \ \ \ }\frac{\theta ^{n-1}}{n}.
\end{equation*}%
For high types%
\begin{equation*}
\int_{\theta ^{n-1}}^{1}pdG_{h}\left( p\right) \text{ \ \ \ \ }=\text{ \ \ \
\ }\left. pG_{h}\left( p\right) \right\vert _{\theta
^{n-1}}^{1}-\int_{\theta ^{n-1}}^{1}G_{h}\left( p\right) dp\text{ \ \ \ \ }=%
\text{ \ \ \ \ }\frac{1-\theta ^{n}}{\left( 1-\theta \right) n}.
\end{equation*}%
The seller collects all these bids, thus%
\begin{eqnarray*}
\pi \left( n\right) &=&\sum_{j=0}^{n}\binom{n}{j}\left( 1-\theta \right)
^{j}\theta ^{n-j}\left[ \left( n-j\right) \frac{\theta ^{n-1}}{n}+j\frac{%
1-\theta ^{n}}{\left( 1-\theta \right) n}\right] \\
&=&\theta ^{n-1}\sum_{j=0}^{n}\binom{n}{j}\left( 1-\theta \right) ^{j}\theta
^{n-j}+\frac{1-\theta ^{n-1}}{\left( 1-\theta \right) n}\sum_{j=0}^{n}\binom{%
n}{j}\left( 1-\theta \right) ^{j}\theta ^{n-j}j
\end{eqnarray*}%
The first term involving the summation sign is equal to 1, as it represents
the sum of the probabilities under $binomial\left( n,1-\theta \right) $. The
second term involving the summation sign is equal to $n\left( 1-\theta
\right) ,$ as it is the expected value under $binomial\left( n,1-\theta
\right) $. Therefore 
\begin{equation*}
\pi \left( n\right) =\theta ^{n-1}\times 1+\frac{1-\theta ^{n-1}}{\left(
1-\theta \right) n}\times n\left( 1-\theta \right) =1
\end{equation*}%
The expected profit of an all pay seller is easier to calculate. We have%
\begin{equation*}
\pi _{ap}\left( 1\right) =r\text{ \ and }\pi _{ap}\left( n\right) =1\text{
for }n\geq 2
\end{equation*}%
hence 
\begin{equation}
\Pi _{ap}=z_{1}\left( x_{h,ap}+x_{l,ap}\right) r+1-z_{0}\left(
x_{h,ap}+x_{l,ap}\right) -z_{1}\left( x_{h,ap}+x_{l,ap}\right) .
\label{Profit Fixed}
\end{equation}%
It is easy to verify that 
\begin{equation*}
\Pi _{ap}=1-z_{0}\left( x_{h,ap}+x_{l,ap}\right)
-x_{h,ap}U_{h,ap}-x_{l,ap}U_{l,ap},
\end{equation*}%
where $U_{h,ap}$ and $U_{l,ap}$ are given by (\ref{Ua}). This completes the
proof. $\blacksquare $

\bigskip

\textbf{Proof of Proposition \ref{prop - main}}. The proof involves two
steps. First we will characterize the outcome where all sellers compete via
all pay auctions. Then we will show that no seller has a profitable
deviation by selecting 1st or 2nd price auctions.

As for the first task, conjecture an outcome where sellers compete with all
pay auctions. Symmetry in buyers' visiting strategies implies that $%
x_{l,a}=\lambda _{l}$ and $x_{h,a}=\lambda _{h};$ so, the total demand at
each store is $\lambda _{l}+\lambda _{h}=\lambda $. When $b>\sigma ,$ the
expected utility of a high type as well as a low type is equal to%
\begin{equation*}
U_{h,ap}=U_{l,ap}=z_{0}(\lambda )(1-r).
\end{equation*}%
A seller solves 
\begin{equation*}
\max_{\lambda }1-z_{0}\left( \lambda \right) -\lambda \Omega ,
\end{equation*}%
which yields $z_{0}\left( \lambda \right) =\Omega .$ It follows that%
\begin{equation*}
z_{0}(\lambda )(1-r)=z_{0}\left( \lambda \right)
\end{equation*}%
implying that the equilibrium reserve price is $r^{\ast }=0$ and the seller
earns 
\begin{equation*}
\Pi =1-z_{0}\left( \lambda \right) -z_{1}\left( \lambda \right) .
\end{equation*}

We now show that a seller cannot do better by switching to a first of second
price format. At all-pay stores we have $U_{h}=U_{l}=\Omega ,$ whereas at
the deviating store we have $U_{h}^{\prime }>U_{l}^{\prime }.$ There are
three scenarios, therefore:\ 

\begin{itemize}
\item The store attracts low types, while high types stay away. This
requires $U_{l}^{\prime }=\Omega >U_{h}^{\prime },$ but contradicts the fact
that $U_{h}^{\prime }>U_{l}^{\prime }.$

\item The store attracts both types. This requires $U_{l}^{\prime }=\Omega $
and $U_{h}^{\prime }=\Omega ,$ but contradicts $U_{h}^{\prime
}>U_{l}^{\prime }$.

\item The store attracts high types, while low types stay away, i.e. $%
U_{h}^{\prime }=\Omega >U_{l}^{\prime }.$ This is feasible, so we focus on
this scenario.
\end{itemize}

The seller solves%
\begin{equation*}
\max_{x}\Pi ^{\prime }=1-z_{0}\left( x\right) -xU_{h}^{\prime }\text{ \ s.t.
\ }U_{h}^{\prime }=\Omega 
\end{equation*}%
The first order condition implies $z_{0}\left( x\right) =\Omega .$ When $%
x_{l}=0,$ under the first-price or second-price formats we have 
\begin{equation*}
U_{h}^{\prime }=z_{0}\left( x\right) \left( 1-r\right) 
\end{equation*}%
Solving $z_{0}\left( x\right) \left( 1-r\right) =z_{0}\left( x\right) $
implies that $r=0.$ Furthermore since $U_{h}^{\prime }=\Omega =z_{0}\left(
\lambda \right) $ implies $x=\lambda .$ Consequently, the seller earns as
much as $\Pi $, which does not present an incentive to deviate; hence the
all pay equilibrium remains. Its uniqueness follows from Proposition \ref%
{prop-no 1st2ndeq'm}. $\blacksquare $

\end{document}